\documentclass[a4paper,10pt,twoside]{cpc-hepnp}

\usepackage{multicol}
\usepackage{graphicx}
\usepackage{booktabs}
\usepackage{amssymb,bm,mathrsfs,bbm,amscd}
\usepackage[tbtags]{amsmath}
\usepackage{lastpage}

\begin{document}



\title{The role of hyperfine mixing in $b\to c$ semileptonic decays of
doubly-heavy baryons\thanks{Supported by DGI and FEDER funds, under contracts
  FIS2008-01143/FIS, FIS2006-03438, FPA2007-65748, CSD2007-00042, by Junta de
  Castilla y Le\'on under contracts SA016A07 and GR12, and by the EU 
  HadronPhysics2 project}}

\author{%
      C. Albertus$^{1,2;1)}$\email{albertus@usal.es}%
\quad E. Hern\'andez$^{1;2)}$\email{gajatee@usal.es}%
\quad J. Nieves$^{2;3)}$\email{jmnieves@ific.uv.es}
}
\maketitle

\address{%
1~(Departamento de F\'{\i}sica Fundamental e IUFFyM, Universidad de Salamanca,
E-37008 Salamanca, Spain)\\
2~(Instituto de  F\'{\i}sica Corpuscular (IFIC), Centro Mixto CSIC-Universidad de
Valencia, Institutos de Investigaci\'on de Paterna,
  Aptd. 22085, E-46071 Valencia, Spain)\\
}

\begin{abstract}
We analyze the effects of hyperfine mixing in  $b\to c\, $ semileptonic decays of doubly heavy 
baryons. We qualitatively confirm the results  by W. Roberts and M. Pervin in Int. J. Mod. Phys. A,
2009, {\bf 24}: 2401-2413, finding that mixing has a great impact on those transitions. However, predictions
 without mixing differ by a factor of 2 and this discrepancy translates to the mixed
case where large differences in decay widths  are observed between the two calculations.
\end{abstract}



\begin{multicols}{2}

\section{Introduction}

According to heavy quark spin symmetry, in the infinite heavy quark mass  limit, one can select
the total spin ($S_h$) of the heavy quark subsystem of a doubly heavy baryon to be
 well defined\cite{jenkins}. This result has 
been used as a basis for the usual classification scheme of doubly heavy baryons. In Table
\ref{tab:clasif} we list the ground state doubly heavy baryons with quantum numbers
$J^\pi=\frac12^+,\,\frac32^+$ that are included in this study.
\end{multicols}
\begin{center}
\tabcaption{ \label{tab:clasif} Quantum numbers and quark content of doubly
heavy baryons }
\vspace{-3mm}
\footnotesize
\begin{tabular*}{170mm}{@{\extracolsep{\fill}}cccc||cccc}\toprule
Baryon & Quark content & $S_h$ & $J^\pi$ &Baryon & Quark content & $S_h$ &
$J^\pi$\vspace{-.1cm}\\ 
&(l=u,d)&&&&&&\\ \hline
$\Xi_{cc}$ & \{c~c\}~l & 1 & 1/2$^+$ & $\Omega_{cc}$ & \{c~c\}~s & 1 & 1/2$^+$ \\ 
$\Xi_{cc}^*$ & \{c~c\}~l & 1 & 3/2$^+$ & $\Omega_{cc}^*$ & \{c~c\}~s & 1 & 3/2$^+$ \\ 
$\Xi_{bb}$ & \{b~b\}~l & 1 & 1/2$^+$ & $\Omega_{bb}$ & \{b~b\}~s & 1 & 1/2$^+$ \\ 
$\Xi_{bb}^*$ & \{b~b\}~l & 1 & 3/2$^+$ & $\Omega_{bb}^*$ & \{b~b\}~s & 1 & 3/2$^+$ \\ 
$\Xi_{bc}$ & \{b~c\}~l & 1 & 1/2$^+$ & $\Omega_{bc}$ & \{b~c\}~s & 1 & 1/2$^+$ \\ 
$\Xi_{bc}^*$ & \{b~c\}~l & 1 & 3/2$^+$ & $\Omega_{bc}^*$ & \{b~c\}~s & 1 & 3/2$^+$ \\ 
$\Xi_{bc}'$ & [b~c]~l & 0 & 1/2$^+$ & $\Omega_{bc}'$ & [b~c]~s & 0 & 1/2$^+$ \\ 
\bottomrule
\end{tabular*}%
\end{center}
\begin{multicols}{2}
Hyperfine interaction between the light and any of the heavy quarks can admix components with
both $S_h=0$ and $S_h=1$ in the wave function. 
The mixing should be very small in the $bb$ and $cc$ sectors  as it will imply higher 
radial excitations or  larger angular momentum. However, mixing  could be 
 particularly important for baryons with $bc$ heavy quark
content where one expects the actual physical $\Xi$ ($\Omega$) states to be admixtures of the $\Xi_{bc},\,\Xi'_{bc}$
($\Omega_{bc},\,\Omega'_{bc}$) ones given in Table~\ref{tab:clasif}.  This mixing gives rise to  small changes 
in the masses but, as suggested in Ref.\cite{pervin1}, it could have a great impact on the
widths of decays involving those states. This has been investigated in Ref.\cite{pervin2} using
harmonic oscillator wave functions as an expanding basis. Here we shall try to confirm their findings using our variational
 wave functions described in Ref.~\cite{albertus} and
 obtained with the use of the AL1 potential of Ref.\cite{semay}. 
 All the  details on the calculations  
  can be found in Ref.~\cite{albertus2} and references therein. 
 \section{Results and discussion}
 \subsection{Masses for unmixed states}
 Our results for the masses  are
given in Table \ref{tab:masses}. We compare them with the results obtained in
Ref.\cite{ebert1} using a relativistic  quark model
 that assumes a light quark-heavy diquark structure, and in the above mentioned Ref.\cite{pervin1} where they use
 a nonrelativistic approach with harmonic oscillator wave functions. The agreement with the
 calculation in Ref.\cite{ebert1} is very good for $\Xi$ baryons whereas for $\Omega$ baryons
 their masses are some $50\sim 90\,$ MeV larger. The masses obtained in Ref.\cite{pervin1} 
 are always
 larger than ours by  $ 50\sim 180\,$MeV. 
On the experimental side the SELEX Collaboration  claimed
evidence for  the $\Xi^+_{cc}$ baryon, in the $\Lambda_c^+K^-\pi^+$ and 
$pD^+K^-$ decay modes, with a mass of
$M_{\Xi^+_{cc}}=3519\pm 1\ \mathrm{MeV/c^2}$~\cite{mattson}, a 100\,MeV smaller than
most theoretical predictions. No other experimental 
collaboration has found  evidence for doubly charmed baryons so far and,   at present,
 the    $\Xi^+_{cc}$ has only a one star status.
 \end{multicols}
\begin{center}
\tabcaption{ \label{tab:masses} Masses ({\rm in}\ MeV) for unmixed states}
\vspace{-3mm}
\footnotesize
\begin{tabular*}{170mm}{@{\extracolsep{\fill}}lccc||lccc}\toprule
& This work&\cite{ebert1}&\cite{pervin1} & &This work&\cite{ebert1}&\cite{pervin1}\\ \hline
$M_{\Xi_{cc}}$      &3613 &3620 & 3676& $M_{\Omega_{cc}}$        & 3712 &3778 &3815\\
$M_{\Xi_{cc}^*}$    &3707 &3727 &3753 & $M_{\Omega_{cc}^*}$        & 3795 & 3872 &3876 \\ 
$M_{ \Xi_{bb} }$    &10198 &10202&10340 & $M_{\Omega_{bb}}$         & 10269 &10359 &10454\\ 
$M_{\Xi_{bb}^*}$     &10237 &10237&10367 & $M_{\Omega_{bb}^*}$        & 10307 &10389&10486\\ 
$M_{\Xi_{bc}}$       &6928 &6933 &7020 &  $M_{\Omega_{bc}}$         & 7013  &7088&7147 \\ 
$M_{\Xi_{bc}'}$      &6958 & 6963  & 7044 &$M_{\Omega_{bc}'}$           & 7038 &7116&7166\\
$M_{\Xi_{bc}^*}$   &6996  &6980 &7078 &$M_{\Omega_{bc}^*}$        & 7075 &7130&7191\\ 
\bottomrule
\end{tabular*}%
\end{center}
\begin{multicols}{2}
\subsection{Decay widths for unmixed states}
Our model to evaluate $b\to c$ semileptonic decays of doubly heavy baryons is described in Ref.\cite{albertus}. We use a 
spectator approximation in which any of the $b$ quarks in the initial state can decay 
into any of the $c$ quarks in the final state. This, together with the right
normalization for baryon states containing two equal heavy quarks, gives an extra factor
$\sqrt2$ in the transition amplitude when compared to the similar $b\to c$ decay in baryons with just one
 heavy quark. 
 
 The results that
we obtain are
shown in Table~\ref{tab:widths} where for comparison we also show the results  in 
Refs.\cite{ebert2,faessler}, obtained within different relativistic approaches, and in 
 the nonrelativistic calculation of
Ref.\cite{pervin2}. Our results are in a global fair agreement with the ones in Ref.\cite{ebert2}.
 As for the other relativistic calculation in Ref.\cite{faessler}, the agreement is fair for
 transitions with a $bc$ baryon in the initial state but there is an approximate
  factor of 2
 discrepancy for transitions with a $bc$ baryon in the final state. The nonrelativistic 
 calculation in Ref.\cite{pervin2} also gives results that are roughly a factor of 2 smaller 
 than ours for all decays. A very interesting
feature of the decay widths shown in Table~\ref{tab:widths} is that they are very different for transitions involving
 $\Xi_{bc}$ or $\Xi'_{bc}$ ($\Omega_{bc}$ or $\Omega'_{bc}$). This means, as suggested in
 Ref.\cite{pervin1}, that mixing in those
 states, provided the admixture coefficients are
 large, can have a great impact on the decay widths.
\end{multicols}
\begin{center}
\tabcaption{ \label{tab:widths} Semileptonic decay widths  $({\rm in}\ 10^{-14}\
{\rm GeV})$ for unmixed states.
We use $|V_{cb}|=0.0413$.  $l=e,\mu$}
\vspace{-3mm}
\footnotesize
\begin{tabular*}{170mm}{@{\extracolsep{\fill}}lcccc||lcccc}\toprule
&\hspace*{-.5cm} This
work\hspace*{-.25cm}&\cite{ebert2}&\hspace*{-.25cm}\cite{faessler}&\hspace*{-.25cm}\cite{pervin2}
&&\hspace*{-.25cm}This
work\hspace*{-.25cm}&\cite{ebert2}&\hspace*{-.25cm}\cite{faessler}&\hspace*{-.25cm}\cite{pervin2}\\\hline
$\Gamma(\Xi_{bb}^*\to\Xi_{bc}'\,l\bar\nu_l)$ &  $1.08$  &$0.82$&
\hspace*{-.25cm}$0.36\pm0.10$&\hspace*{-.25cm}--
&$\Gamma(\Omega_{bb}^*\to\Omega_{bc}'\,l\bar\nu_l)$ & $1.14$
&$0.85$&\hspace*{-.25cm}$0.42\pm0.14$&\hspace*{-.25cm}-- \\ 
$\Gamma(\Xi_{bb}^*\to\Xi_{bc}\,l\bar\nu_l)$
&$0.36$&$0.28$&\hspace*{-.25cm}$0.14\pm0.04$&\hspace*{-.25cm}--
&$\Gamma(\Omega_{bb}^*\to\Omega_{bc}\,l\bar\nu_l)$
&$0.38$&$0.29$&\hspace*{-.25cm}$0.15\pm0.05$&\hspace*{-.25cm}--\\ 
$\Gamma(\Xi_{bb}\to\Xi_{bc}'\,l\bar\nu_l)$ &  $1.09$ &$0.82$&\hspace*{-.25cm}$0.43\pm0.12$&
\hspace*{-.25cm}$0.41$
&$\Gamma(\Omega_{bb}\to\Omega_{bc}'\,l\bar\nu_l)$ & $1.16$
&$0.83$&\hspace*{-.25cm}$0.48\pm0.12$&\hspace*{-.25cm}$0.51$ \\ 
$\Gamma(\Xi_{bb}\to\Xi_{bc}\,l\bar\nu_l)$ 
&$2.00$&$1.63$&\hspace*{-.25cm}$0.80\pm0.30$&\hspace*{-.25cm}$0.69$
&$\Gamma(\Omega_{bb}\to\Omega_{bc}\,l\bar\nu_l)$  &
$2.15$&$1.70$&\hspace*{-.25cm}$0.86\pm0.32$&\hspace*{-.25cm}$0.92$\\ 
$\Gamma(\Xi_{bc}'\to\Xi_{cc}\,l\bar\nu_l)$ & 
$1.36$&$0.88$&\hspace*{-.25cm}$1.10\pm0.32$&\hspace*{-.25cm}--
&$\Gamma(\Omega_{bc}'\to\Omega_{cc}\,l\bar\nu_l)$ &
$1.36$&$0.95$&\hspace*{-.25cm}$0.98\pm0.28$&\hspace*{-.25cm}-- \\ 
$\Gamma(\Xi_{bc}\to\Xi_{cc}\,l\bar\nu_l)$  &$ 2.57
$&$2.30$&\hspace*{-.25cm}$2.10\pm0.70$&\hspace*{-.25cm}$1.38$
&$\Gamma(\Omega_{bc}\to\Omega_{cc}\,l\bar\nu_l)$  &
$2.58$&$2.48$&\hspace*{-.25cm}$1.88\pm0.62$&\hspace*{-.25cm}$1.54$\\ 
$\Gamma(\Xi_{bc}'\to\Xi_{cc}^*\,l\bar\nu_l)$ & 
$2.35$&$1.70$&\hspace*{-.25cm}$2.01\pm0.62$&\hspace*{-.25cm}-- 
&$\Gamma(\Omega_{bc}'\to\Omega_{cc}^*\,l\bar\nu_l)$ & $2.35$
&$1.83$&\hspace*{-.25cm}$1.93\pm0.60$&\hspace*{-.25cm}--\\ 
$\Gamma(\Xi_{bc}\to\Xi_{cc}^*\,l\bar\nu_l)$  &$ 0.75 $ &$0.72$&\hspace*{-.25cm}$0.64\pm0.19$&\hspace*{-.25cm}$0.52$
&$\Gamma(\Omega_{bc}\to\Omega_{cc}^*\,l\bar\nu_l)$ 
&$0.76$&$0.74$&\hspace*{-.25cm}$0.62\pm0.19$&\hspace*{-.25cm}$0.56$\\
\bottomrule
\end{tabular*}%
\end{center}
\begin{multicols}{2}\vspace{1cm}

\subsection{Results with mixing}

We obtain the mixed  ${bc}$ states by diagonalization of the corresponding mass
matrices. In our calculation 
the mixed states and masses are given by
\begin{eqnarray}
\label{eq:mix}
&&\hspace*{-.9cm}\Xi\,_{bc}^{(1)}=
\hspace{.3cm}0.902\,\Xi\,'_{bc}+0.431\,\Xi\,_{bc}\,,\ 
M_{\Xi\,_{bc}^{(1)}}=6967\,{\rm MeV},\nonumber\\
&&\hspace*{-.9cm}\Xi\,_{bc}^{(2)}= -0.431\,\Xi\,'_{bc}+0.902\,\Xi\,_{bc}\, ,\ 
M_{\Xi\,_{bc}^{(2)}}= 6919\,{\rm MeV},\nonumber\\
&&\hspace*{-.9cm}\Omega\,_{bc}^{(1)}=
\hspace{.25cm}0.899\,\Omega\,'_{bc}+0.437
\,\Omega\,_{bc}\,,\ 
M_{\Omega\,_{bc}^{(1)}}=7046\,{\rm MeV},\nonumber\\
&&\hspace*{-.9cm}\Omega\,_{bc}^{(2)}=
-0.437\,\Omega\,'_{bc}+0.899\,\Omega\,_{bc}\, ,\ 
M_{\Omega\,_{bc}^{(2)}}= 7005\,{\rm MeV}.\nonumber\\
\end{eqnarray}
By comparison to the unmixed results shown in Table~\ref{tab:masses}, we see the masses change but very
 little when mixing is taken in to account. However,  as shown in 
 Eq.(\ref{eq:mix}),  the admixture is important and it can affect the decay widths.

Note that these mixed states are close to the states (in the what follows $B\equiv\Xi,\,\Omega$)
\begin{eqnarray}
&&\hspace*{-1.1cm}B^{(1)}_{bc}\approx\left(|qc;1\rangle\otimes|b;\frac12\rangle\right)^{J=1/2}\equiv
\frac{\sqrt3}{2}B'_{bc}
+\frac{1}{2}B_{bc},\nonumber\\
&&\hspace*{-1.1cm}B^{(2)}_{bc}\approx\left(|qc;0\rangle\otimes|b;\frac12\rangle\right)^{J=1/2}\equiv
-\frac{1}{2}B'_{bc}+\frac{\sqrt3}{2}B_{bc},
\end{eqnarray}
in which the light and the $c$ quark couple to well defined spin 1 or 0. 

\end{multicols}
\begin{center}
\tabcaption{ \label{tab:newwidths} Semileptonic decay widths  $({\rm in}\ 10^{-14}\ {\rm GeV})$ for mixed states.
We use $|V_{cb}|=0.0413$.  $l=e,\mu$ }
\vspace{-3mm}
\footnotesize
\begin{tabular*}{170mm}{@{\extracolsep{\fill}}lcc||lcc}\toprule
  &This work&\cite{pervin2}&&This work&\cite{pervin2}\\\hline
$\Gamma(\Xi_{bb}^*\to\Xi^{(1)}_{bc}\,l\bar\nu_l)$ &  $0.47$& --  
&$\Gamma(\Omega_{bb}^*\to\Omega^{(1)}_{bc}\,l\bar\nu_l)$ & $0.48$&--  \\ 
$\Gamma(\Xi_{bb}^*\to\Xi^{(2)}_{bc}\,l\bar\nu_l)$\ &$0.99$&--
&$\Gamma(\Omega_{bb}^*\to\Omega^{(2)}_{bc}\,l\bar\nu_l)$\ &$1.06$&--\\ 
$\Gamma(\Xi_{bb}\to\Xi^{(1)}_{bc}\,l\bar\nu_l)$ &  $2.21$& $0.95$ 
&$\Gamma(\Omega_{bb}\to\Omega^{(1)}_{bc}\,l\bar\nu_l)$ & $2.36$& $0.99$  \\ 
$\Gamma(\Xi_{bb}\to\Xi^{(2)}_{bc}\,l\bar\nu_l)$  &$0.85$& $0.33$
&$\Gamma(\Omega_{bb}\to\Omega^{(2)}_{bc}\,l\bar\nu_l)$  & $0.91$& $0.30$\\ 
$\Gamma(\Xi^{(1)}_{bc}\to\Xi_{cc}\,l\bar\nu_l)$ &  $0.38$& --
&$\Gamma(\Omega^{(1)}_{bc}\to\Omega_{cc}\,l\bar\nu_l)$ & $0.37$& -- \\ 
$\Gamma(\Xi^{(2)}_{bc}\to\Xi_{cc}\,l\bar\nu_l)$  &$ 3.51$&$ 1.92 $
&$\Gamma(\Omega^{(2)}_{bc}\to\Omega_{cc}\,l\bar\nu_l)$  & $3.52$& $1.99$\\ 
$\Gamma(\Xi^{(1)}_{bc}\to\Xi_{cc}^*\,l\bar\nu_l)$ &  $3.14$&--
&$\Gamma(\Omega^{(1)}_{bc}\to\Omega_{cc}^*\,l\bar\nu_l)$ & $3.14$&-- \\ 
$\Gamma(\Xi^{(2)}_{bc}\to\Xi_{cc}^*\,l\bar\nu_l)$  &$0.017$& $0.026$
&$\Gamma(\Omega^{(2)}_{bc}\to\Omega_{cc}^*\,l\bar\nu_l)$  &$0.014$&$0.013$\\
\bottomrule
\end{tabular*}%
\end{center}
\begin{multicols}{2}
The new decay widths involving  the mixed states $\Xi\,_{bc}^{(1)},\,\Xi\,_{bc}^{(2)}$ and 
$\Omega\,_{bc}^{(1)},\,\Omega\,_{bc}^{(2)}$ are now given in Table~\ref{tab:newwidths}. We see
rather big changes from the values in Table~\ref{tab:widths} where unmixed states were used. 
Special attention  deserves the
$B^{(2)}_{bc}\to B^*_{cc}$ transitions where the width reduces by a large factor of $44$ (54) for the 
$\Xi^{(2)}_{bc}\to\Xi^*_{cc}$ ($\Omega^{(2)}_{bc}\to \Omega^*_{cc}$) decay compared to
the unmixed case. This can be  easily  understood by taking into account that
$B^{(2)}_{bc}\approx\left(|qc;0\rangle\otimes|b;\frac12\rangle\right)^{J=1/2}$. In the latter state the light and 
$c$ quarks are coupled to spin 0, whereas in the $B^*_{cc}$ the light and any of the $c$ quarks are in a
relative spin 1 state. In any spectator calculation, as the ones here and in Ref.\cite{pervin2}, the
amplitude for the $\left(|qc;0\rangle\otimes|b;\frac12\rangle\right)^{J=1/2}\to B^*_{cc}$ 
transition cancels due to the orthogonality of the different spin states of the spectator quarks in the initial and
final baryons. The fact that $B^{(2)}_{bc}$ slightly deviates from$ \left(|qc;0\rangle\otimes|b;\frac12\rangle\right)^{J=1/2}$ 
produces a non zero, but small, decay width.
\section{Conclusions}
We qualitatively confirm the findings in Ref.\cite{pervin2} as to the relevance of hyperfine mixing in 
$b\to c$ semileptonic decays of doubly heavy baryons. On the other hand the absolute
predictions are quite different. This is a reflection of the approximate factor of 2 difference we already found 
in Table~\ref{tab:widths} for unmixed states.
 
 \acknowledgments{C. A.  acknowledges a contract supported by PIE-CSIC 200850I238
during his stay at IFIC.}\vspace{-.2cm}\\

\end{multicols}

\clearpage

\end{document}